\documentclass[a4paper]{PoS}
\title{Studies of the muon-induced 
neutron background in LSM:
detector concept and status of the installation }
\ShortTitle{Studies of the muon-induced neutrons in LSM}
\author{\speaker{V.~Yu.~Kozlov}
(for the Edelweiss collaboration)\\
Forschungszentrum Karlsruhe, Institut f\"ur Kernphysik, Postfach 3640, 76021 Karlsruhe, Germany \\
E-mail: \email{Valentin.Kozlov@ik.fzk.de}}

\abstract{A good particle candidate for Cold Dark Matter (CDM) is the supersymmetric neutralino or more generally a weakly interacting massive particle (WIMP). The expected interaction rate of WIMPs with the detector medium in the direct detection experiments is below 0.01 events/kg/day. This makes a good knowledge of the background conditions highly important, especially with ever increasing sensitivity of the detectors. One of the background components is related to cosmic muons and in particular to muon-induced neutrons. Detailed studies carried out by the Edelweiss collaboration in this respect are presented. This activity includes GEANT4 simulations with full event topology as well as a dedicated measurement with a new neutron counter installed in the fall of 2008 in LSM (Laboratoire Souterrain de Modane, France). This counter is incorporated into the existing muon veto system thus allowing to monitor neutrons in coincidence with the incoming muons.\\

}

\FullConference{Identification of dark matter 2008\\
		 August 18-22, 2008\\
		 Stockholm, Sweden}

\begin{document}
\section{Introduction}
The Edelweiss experiment searches for the WIMP candidates of the Dark Matter. The set-up is located in LSM in the French Alps which provides a shielding factor of $\sim$4800~m.w.e. The detection principle is based on measuring energy of the recoil nucleus originating from the WIMP elastic scattering. Bolometers of pure natural Ge are used both as the detectors and the target material. These detectors cooled down to about 20~mK allow to measure simultaneously heat and ionization signals. Due to the quenching of the ionization signal present for nuclear recoils one achieves a very high discrimination of $\beta$ and $\gamma$ background from the recoil candidates \cite{defay08}. However, neutrons coming from the natural radioactivity or induced by the remaining muons can still mimic the nuclear recoil signal of WIMP events and thus can not be discriminated in the same way as $\beta$'s and $\gamma$'s. Therefore, this type of background requires special careful investigation. The knowledge of it also becomes highly important in view of large 1-tonne scale experiments like EURECA \cite{kraus08}.

\section{Neutron background}
For kinematic reasons, not every neutron can mimic the WIMP nuclear recoil event but only those who have an energy of 0.5\--10~MeV when they reach the Ge bolometers. Such neutrons appearing due to natural radioactivity in the surrounding (e.g. U/Th contamination) can be avoided by using a passive hydrogen-rich moderator (50~cm polyethylene shield in case of Edelweiss) and in addition by radiopurity selection of materials to be used. Monitoring of the ambient neutron flux in proximity of the Edelweiss experimental set-up is performed with the help of $^3$He gas detectors. This measurement yields a flux of about $2\cdot10^{-6}$~n/cm$^2$/day \cite{yakushev08} which is in good agreement with the previously measured value \cite{fiorucci07}. Another part of the neutron background is caused by muon interactions in the rock and in fact everywhere in the set-up (especially in high-Z materials such as the gamma shield based on lead). High energy neutrons (well above 10~MeV) created in such deep inelastic scattering (DIS) processes further lead to the production of secondary neutrons with energies below 10~MeV. The effect of this $\mu-$induced neutron component is commonly reduced by tagging the original muons. In Edelweiss experiment the plastic scintillator modules covering the full bolometer set-up act as the muon veto \cite{chantelauze07}. Full simulations of the Edelweiss set-up including the muon veto were performed in GEANT4 in order to estimate the influence of muons for the Dark Matter search. These simulations involve muon generation to reproduce the muon flux specific for LSM and allow to get complete event topology \cite{horn07}. It was shown then that muons which miss the veto can still induce some neutrons reaching the bolometers and giving rise to WIMP-like events not vetoed by the muon system. To verify these simulations one has to normalize them to the experimental data, i.e. one needs explicit $\mu-$induced neutron measurements. One way to achieve this is to check the rate of events which are in coincidence between muon veto and the bolometers. This rate currently measured in Edelweiss is about 0.03~events/kg/day and it is reasonably well reproduced in the simulations. However, the rareness of these coincidence events makes it hard to get enough statistics to draw a reliable conclusion. A dedicated detector based on liquid scintillator was thus designed and installed in 2008 in LSM.

\section{A detector for muon-induced neutrons}
\label{sec:nc-detector}

\begin{figure}[!ht]
	\centering
		\includegraphics[width=0.72\textwidth]{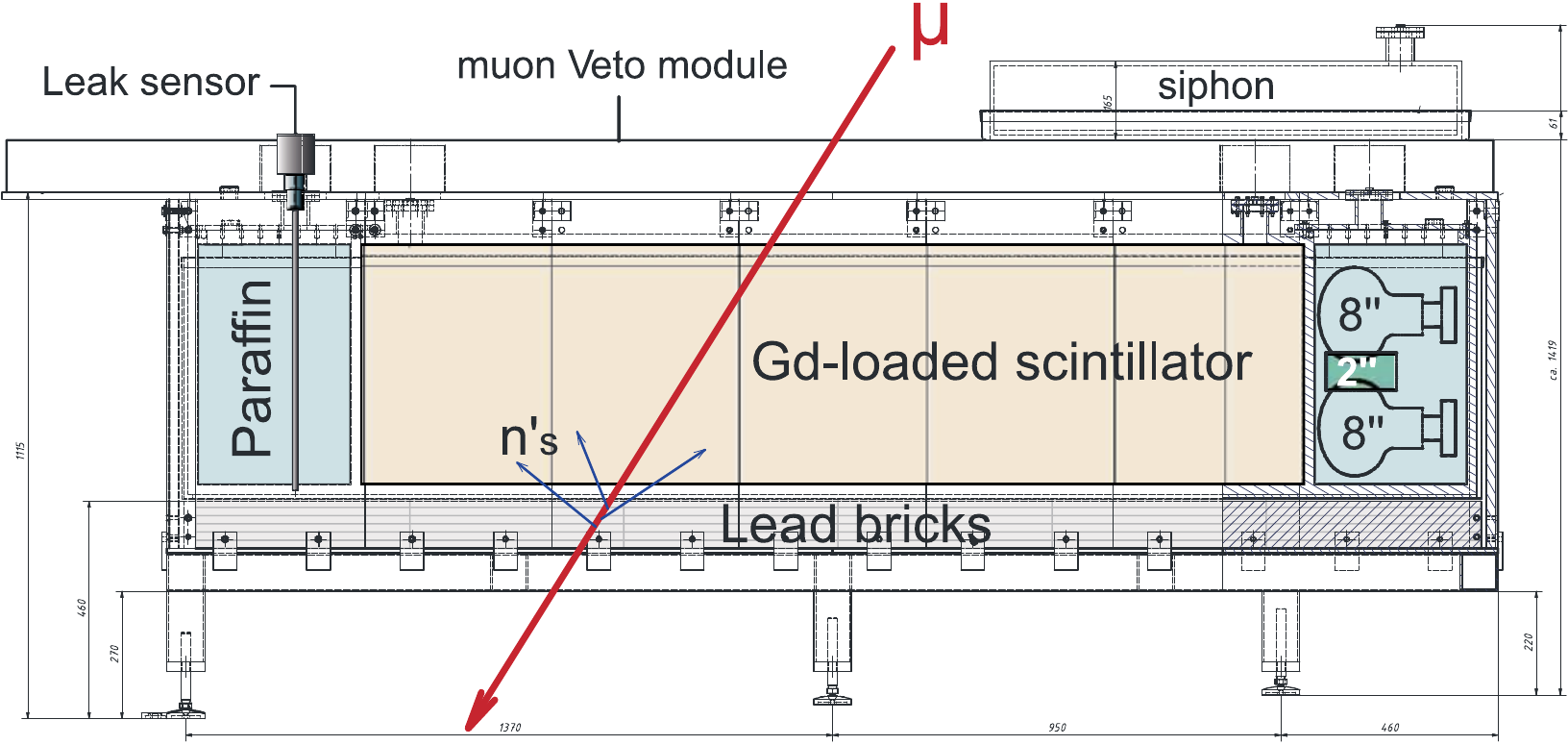}
	\caption{General scheme of the neutron counter (side view): 1-tonne of Gd-loaded liquid scintillator viewed by 16 PMTs of 8-inch and 6 PMTs of 2-inch diameter. A layer of lead bricks below the liquid scintillator volume acts as an effective target for muons and high energy neutrons.}
	\label{fig:nc-detector}	
\end{figure}

The measuring principle of $\mu$-induced neutrons is based on registering thermalized neutrons in coincidence with the incoming muon or by detecting a multiple neutron event (secondary neutrons produced in a $\mu$-induced particle showers). To efficiently observe neutrons, a liquid scintillator of 1~m$^3$ volume (50x100x200~cm$^3$) loaded with Gadolinium (St.~Gobain Bicron BC525) is used as a core of the detector. The neutron capture process on Gd results in several gammas with 8~MeV sum energy. The scintillator volume is viewed from each of the two module ends by 8 photomultiplier tubes (PMT) of 8-inch diameter (Fig.~\ref{fig:nc-detector}). These PMTs are optimized to register the light produced after the neutron capture. In addition, the system is equipped with 6 smaller PMTs (2-inch diameter) to register muons crossing the neutron detector (these muons create much more light and thus the 8-inch PMTs will saturate). The scintillator and PMTs are placed in one plexiglass container divided into three parts: the central one for the scintillator itself and two side ones filled with paraffin in which the PMTs are immersed. This plexiglass chamber is then placed in an aluminum vessel as secondary safety container. Finally, the system is surrounded by iron plates to reflect a fraction of neutrons back to the scintillator. In order to enhance the neutron production (up to a factor of 10 comparing to rock) a 10~cm thick layer of lead bricks is put underneath the detector. On top of the counter, a plastic scintillator module (same type as the muon veto of Edelweiss) is installed. The complete system is positioned right near the western wall of the Edelweiss muon veto. Based on the currently measured muon flux in the lab, the expected count rate of muon-induced neutrons is about few counts per day. 

The GEANT4 simulations mentioned above were extended to optimize the neutron set-up before going for construction. Additionally, a smaller prototype (25x25x250~cm$^3$) was built beforehand in Karlsruhe in order to test mechanical properties, handling of liquids and gas as well as to study light collection, PMTs and overall performance. This prototype also allowed to develop a LED system to monitor over time the light properties of the scintillator and stability of PMTs. There are in total 8 LEDs ($\lambda=$425~nm) placed at different positions. These LEDs are operated via VME-based PC commands and regularly fired one by one. The data from the groups of opposite PMTs are then analyzed. 

\begin{figure}
	\centering
		\includegraphics[width=0.62\textwidth]{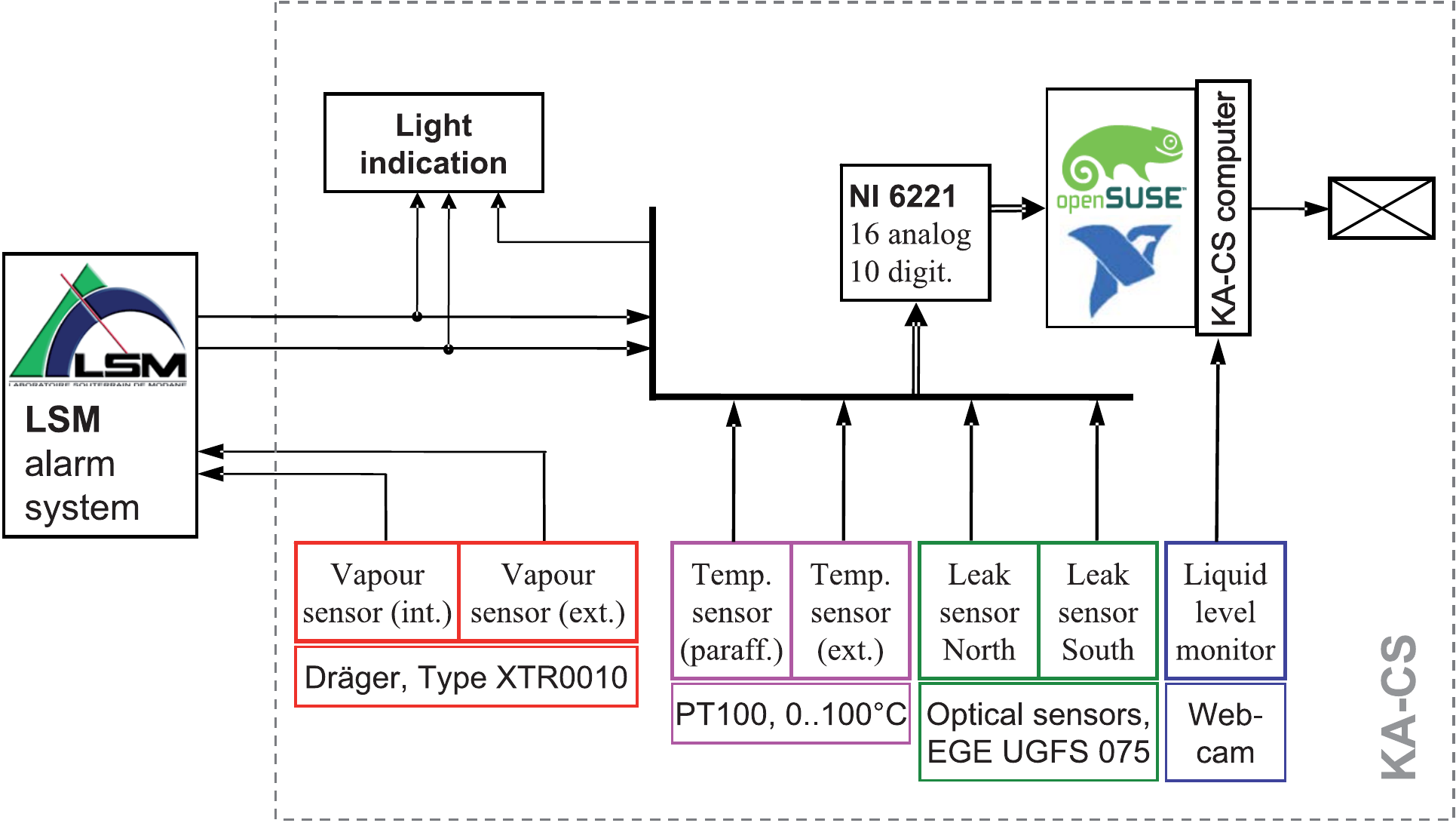}
	\caption{Principle scheme of KA-CS system for safety monitoring: sensors are continuously read via NI-6221 DAQ card by the LabVIEW program installed on Linux computer.}
	\label{fig:nc-ka-cs}
\end{figure}

The neutron counter is also equipped with safety sensors because of the pseudocumene based scintillator. This includes vapor sensors to check the internal and surrounding atmosphere, two leak sensors in the aluminum vessel, one temperature meter immersed in the paraffin volume and one outside of the counter. Signals of the vapor sensors are incorporated into LSM safety system which takes care of an alarm activation in the lab. One can as well monitor these sensors using the LabVIEW$^{\tt{TM}}$-based program (\underline{Ka}rlsruhe \underline{C}ontrol of \underline{S}afety or KA-CS) installed on Linux computer (SuSE~10.3) (Fig.~\ref{fig:nc-ka-cs}). This software notifies users by email in case of an alarm due to a failure or passing of specified thresholds.


The neutron detector described was successfully installed in LSM in September 2008 and as for the time of writing, it is under intensive commissioning.

\section{Conclusion}
Improved sensitivity of Dark Matter search experiments requires much better knowledge of the background conditions. Activity of the Edelweiss collaboration concerning the neutron background studies was presented, in particular the new detector for the $\mu$-induced neutrons was described. 

\section{Acknowledgements}
This work is in part supported by the German Research Foundation (DFG) through the Trans\-regional Collaborative Research Center SFB-TR27 as well as by the EU contract RII3-CT-2004-506222.

\end{document}